\magnification 1200
\centerline {\bf Macrostatistics and Fluctuating Hydrodynamics}
\vskip 0.5cm
\centerline {by Geoffrey L. Sewell\footnote*{e-mail: g.l.sewell@qmul.ac.uk}}
\vskip 0.5cm
\centerline {Department of Physics, Queen Mary University of London}
\vskip 0.2cm
\centerline {Mile End Road, London E1 4 NS, UK}
\vskip 1cm
\centerline {\bf Abstract}
\vskip 0.5cm
We extend our earlier macrostatistical treatment of hydrodynamical fluctuations about 
nonequilibrium steady states to viscous fluids. Since the scale dependence of the 
Navier-Stokes equations precludes the applicability of any infinite scale 
(hydrodynamical) limit, this has to based on the generic model of a large but finite 
system, rather than an infinite one. On this basis, together with the assumption of 
Onsager\rq s regression hypothesis and conditions of  local equilibrium and 
chaoticity, we show that  the hydrodynamical fluctuations of  a reservoir driven fluid 
about a nonequilibrium steady state execute a Gaussian Markov process that 
constitutes a mathematical structure for a generalised version of Landau\rq s 
fluctuating hydrodynamics and generically carries long range spatial correlations. 
\vfill\eject
\centerline {\bf 1. Introduction}
\vskip 0.3cm
The theory of  fluctuations is a key area of statistical physics, which is essential to 
both equilibrium [1] and nonequilibrium thermodynamics [2]. Further, Landau [3] 
introduced fluctuation theory into fluid dynamics by adding white noise terms to the 
Navier-Stokes equations and employing the fluctuation-dissipation theorem to relate 
their intensities to the viscosity and thermal conductivity coefficients.
\vskip 0.2cm
In relatively recent works [4, 5], we have presented a macrostatistical approach to the 
theory of  hydrodynamical fluctuations of reservoir driven quantum systems about 
nonequilibrium steady states. This was designed to relate the stochastic properties of 
the fluctuations to the phenomenological continuum mechanics of these systems on 
the basis of  general arguments centred on the hydrodynamical observables, subject to 
assumptions of  macroscopic classicality, local equilibrium, chaoticity and a 
generalised version of Onsager\rq s regression hypothesis\footnote*{In the present 
setting, this signifies essentially that small deviations of the hydrodynamical 
observables from their values in a nonequilibrium steady state evolve according to the 
same dynamical law whether this deviation arises from a spontaneous fluctuation or 
from a small external perturbation.} [2], all pertaining to states that may be far from 
global thermal equilibrium. On this basis, it was shown that, in a certain large scale 
limit, the fluctuations of the hydrodynamical variables executed a classical 
macroscopic stochastic process, whose parameters were expressed in terms of purely 
phenemenological quantities, namely thermodynamical variables and transport 
coefficients, the underlying quantum mechanics of the system being buried in the 
forms of these variables.  Among the results that ensued from a treatment of this 
stochastic process [4, 5] were a non-linear generalisation of Onsager\rq s reciprocity 
relations and a proof that the spatial correlations of the hydrodynamical variables are 
generically of long range. However, as it stands, the theory of those works is limited 
to systems whose phenomenological evolutions are scale 
invariant\footnote{**}{Specifically, those evolutions had to be invariant under space-
time scale transformations of the form $x{\rightarrow}{\lambda}x, \ 
t{\rightarrow}{\lambda}^{2}t.$} and is therefore not applicable to viscous fluids, as 
described by the Navier-Stokes (NS) equations, since these are scale dependent.  
\vskip 0.2cm
The object of the present article is to extend our macrostatistical treatment to viscous 
Navier-Stokes fluids and thereby to provide a mathematical structure for a generalised 
form of Landau\rq s picture of hydrodynamical fluctuations and to establish that, 
generically, their spatial correlations in nonequilibrium steady states are of long 
range. However, the scale dependence of the NS equations precludes the applicability 
of a hydrodynamical limit\footnote{***}{This is a limit in which the space and time 
scales for the phenomenological description becomes infinite and the macroscopic 
equations of motion become exact.}, which was basic to the methodology of the 
previous works. Consequently, the theory has to be based on the generic model of a 
large, but finite, system, rather than an infinite one. 
\vskip 0.2cm
Our treatment here is based on a classical macrostatistical model, whose presumed 
relationship to an underlying quantum mechanics is only briefly indicated in the 
concluding Section. Thus, the model comprises a continuous distribution of matter 
that is confined to a bounded spatial region ${\Omega}$ and coupled to an array of 
reservoirs at its boundary. We assume that the scales of mass, distance and time 
whose units represent characteristic values of these variables are macroscopic: thus, 
for example, the constants ${\hbar}$ and $k_{B}$ of Planck and Boltzmann are 
extremely small on these scales\footnote*{For example, if SI units and degrees 
Kelvin are appropriate, ${\hbar}$ and $k_{B}$ are of the order of $10^{-34}$ and 
$10^{-23}$, respectively.}. We take the hydrodynamical observables to be the 
position and time dependent densities of energy, mass and momentum, and we 
assume that their phenomenological evolution is given by the NS equations. We then 
construct the model of the hydrodynamical fluctuations about a nonequilibrium steady 
state on the basis of assumptions of  the Onsager regression hypothesis, local 
equilibrium and chaoticity. Although no hydrodynamical limit is available here, we 
exploit the fact that Boltzmann\rq s constant is extremely small on the employed 
macroscopic scaling: specifically, we pass to a limit in which $k_{B}{\rightarrow}0$ 
in our formulation of local equilibrium conditions via Einstein\rq s relation, $P={\rm 
const.}{\rm exp}(S/k_{B})$, between the equilibrium probability distribution of the 
macroscopic variables and the entropy function $S$. On this basis we obtain a 
generalisation of Landau\rq s picture, wherein the hydrodynamical fluctuations 
execute a Gaussian Markov process whose parameters are completely determined by 
macroscopic variables involved in the phenomenological thermodynamic and 
hydrodynamic pictures of the system: the underlying quantum mechanics is assumed 
to be buried in the forms of these variables as functions of the control parameters. 
Futhermore, we show that the spatial correlations of the hydrodynamical variables are 
generically of long range.  
\vskip 0.2cm
We present our treatment as follows. In Section 2 we formulate the thermodynamics
 and hydrodynamics of the model in purely phenomenological terms. In Section 3 we 
construct the stochastic process executed by the hydrodynamical fluctuations, subject 
to the assumptions specified in the previous paragraph. In Section 4 we prove that the 
spatial correlations of these fluctuations are generically of long range. We conclude in 
Section 5 with some brief comments on the basis of the model and on its 
presumed relationship to its underlying quantum mechanics.  There are two 
Appendices: the first is devoted to a calculation leading to a key formula, the second 
to a proof of a lemma.
\vskip 0.3cm
{\it Note on distributions.} As we shall represent the hydrodynamical fluctuations by 
distributions, in the sense of L. Schwartz [6], we now specify our notations for 
these.We denote by ${\cal D}({\Omega}), \ {\cal D}_{V}({\Omega})$ and 
${\cal D}_{T}({\Omega})$ the Schwartz spaces of real valued, infinitely 
differentiable scalar, vector and second order tensor valued functions on the bounded 
open region ${\Omega}$ with support in that region. These spaces are reflexive and 
their duals are distributions, which we denote by ${\cal D}^{\prime}({\Omega}), \ 
{\cal D}_{V}^{\prime}({\Omega})$ and ${\cal D}_{T}^{\prime}({\Omega})$, 
respectively. We define ${\tilde {\cal D}}({\Omega})$ and ${\check {\cal 
D}}({\Omega})$ to be the Cartesian products ${\cal D}({\Omega}){\times}{\cal 
D}({\Omega}){\times}{\cal D}_{V}({\Omega})$ and  ${\cal 
D}_{V}({\Omega}){\times}{\cal D}_{V}({\Omega}){\times}{\cal 
D}_{T}({\Omega})$, respectively, and we denote their duals by ${\tilde {\cal 
D}}^{\prime}({\Omega})$ and ${\check {\cal D}}^{\prime}({\Omega})$, 
repectively. Further, if ${\alpha}$ and ${\psi}$ are elements of a ${\cal D}$-class 
space and its dual, respectively, then ${\psi}({\alpha})$ will sometimes be denoted by 
${\langle}{\psi},{\alpha}{\rangle}$. We also employ angular brackets to denote the 
$L^{2}$ inner product ${\langle}{\alpha},{\alpha}^{\prime}{\rangle}$ between 
pairs of elements ${\alpha}$ and ${\alpha}^{\prime}$ of the same ${\cal 
D}({\Omega})$ space. Evidently, these two uses of the angular brackets are mutually 
consistent since any ${\cal D}$ space is a subset of its dual. To avoid ambiguity in the 
definition of scalar products of tensor valued functions, we define that of elements 
${\alpha} \ (={\lbrace}{\alpha}_{ij}{\vert}i,j=(1,2, . \ .,d){\rbrace}$) and 
${\alpha}^{\prime} \ (={\lbrace}{\alpha}_{ij}^{\prime}{\vert}i,j=(1,2, . \ 
.,d){\rbrace})$ of ${\cal D}_{T}({\Omega})$ to be
$${\langle}{\alpha},{\alpha}^{\prime}{\rangle}:=
\int_{\Omega}dx{\alpha}_{ij}(x){\alpha}_{ji}^{\prime}(x),\eqno(1.1)$$
where, as elsewhere in this article, the repeated index summation convention has been 
employed. 
 \vskip 0.5cm
\centerline {\bf 2. The Phenomenological Picture}
\vskip 0.3cm	
We assume that the fluid is a macroscopic system, ${\Sigma}$, that occupies a fixed 
bounded open connected region ${\Omega}$ of a $d$-dimensional Euclidean space 
$X$, which constitutes the laboratory reference frame. We assume that ${\Sigma}$ is 
in contact at its boundary, ${\partial}{\Omega}$,with an array, ${\cal R}$, of 
reservoirs, and that these determine the boundary conditions for the flow of the 
system, in a way that we shall specify in Sec. 2.2. We further assume that its 
dynamics is Galilei covariant. We employ a continuum model for ${\Sigma}$, which 
we formulate on macroscopic scales wherein the magnitudes of its energy, mass and 
volume are of the order of unity: for simplicity, we take its volume to be unity. 
\vskip 0.5cm
\centerline {\bf 2.1. The Thermodynamic Potentials}
\vskip 0.3cm
We assume that, at equilibrium, ${\Sigma}$ is at rest, that its densities, $e_{0}$ and 
${\rho}_{0}$, of energy and mass are spatially uniform and that its entropy density is 
a function, $s_{0}$, of these variables. Thus, as the volume of ${\Omega}$ is unity, 
$e_{0}, \ {\rho}_{0}$ and $s_{0}(e_{0},{\rho}_{0})$ are also the total energy, mass 
and entropy, respectively, of ${\Sigma}$ and satisfy the fundamental formula
$$ds_{0}={\beta}(de_{0}-{\mu}d{\rho}_{0}),\eqno(2.1)$$
where ${\beta}^{-1}$ is the temperature, in degrees Kelvin, say, {\it not} in units of 
$k_{B}$, and ${\mu}$ is the chemical potential, as related to the mass\footnote*{We 
relate the chemical potential to mass, by Eq. (2.3), rather than particle number, since 
the continuum model does not involve any concept of the latter.}. Thus
$${\beta}=\bigl({{\partial}s_{0}\over {\partial}e_{0}}\bigr)_{{\vert}{\rho}_{0}}
\eqno(2.2)$$
and
$${\mu}=-{\beta}^{-1}\bigl({{\partial}s_{0}\over 
{\partial}{\rho}_{0}}\bigr)_{{\vert}{e}_{0}}.\eqno(2.3)$$ 
The pressure is then
$$p={\beta}^{-1}s_{0}(e_{0},{\rho}_{0})-e_{0}+{\mu}{\rho}_{0}.\eqno(2.4)$$
and the heat function (enthalpy density) is
$${\varepsilon}=e_{0}+p.\eqno(2.5)$$ 
\vskip 0.2cm
Turning now to the nonequilibrium situation, the densities of energy, mass and 
momentum are generally non-uniform. We assume that they are locally conserved and 
we denote their local densities by $e, \ {\rho}$ and $j$, respectively. These are 
functions of position $x \ ({\in}{\Omega})$ and time $t \ ({\in}{\bf R})$. The local 
drift velocity is defined to be
$$u=j/{\rho}.\eqno(2.6)$$
We denote the components of $j$ and $u$, relative to some chosen coordinate system, 
by $(j_{1},. \ .,j_{d})$ and $(u_{1},. .,u_{d})$, respectively.
\vskip 0.2cm
We define a local rest frame for the point $x$ to be one that moves with velocity 
$u(x)$ relative to the laboratory frame. It therefore follows from Galilei covariance 
that the local energy and mass densities relative to this frame are 
$$e_{0}=e-{1\over 2}{\rho}u^{2}=e-{j^{2}\over 2{\rho}} \ {\rm and} \ 
{\rho}_{0}={\rho}.\eqno(2.7)$$	
We assume that, even in a nonequilibrium state, the system is in local equilibrium in 
the phenomenological sense\footnote*{A further, macrostatistical kind of local 
equilibrium will be assumed in Section 3.3} that its local enthalpy density 
${\varepsilon}$ is still given by Eq. (2.5) and its entropy density at a point $x$ is the 
same function, $s_{0}$, of the energy and mass densities relative to a local rest frame 
as that governing the equilibrium entropy density relative to the laboratory frame. 
Thus, by the Galilei invariance of entropy 
[7 ], its density at the point $x$, relative to the laboratory frame, is the function $s$ of 
$e(x), \ {\rho}(x)$ and $j(x)$ given by the formula 
$$s(e,{\rho},j)=s_{0}(e_{0},{\rho}_{0}),\eqno(2.8)$$
\vskip 0.2cm
We now compactify the notation by denoting the triple $(e,{\rho},j)$ by ${\phi}$. 
Thus, defining ${\nu}:=(d+2), \ {\phi}$ is the ${\nu}$-component variable 
$({\phi}_{1},. \ .,{\phi}_{\nu}):=(e,{\rho},j_{1},. \ .,j_{d})$ and $s$ is a function of 
${\phi}$. Its derivative $s^{\prime}({\phi}) \bigl( :=({\partial}s/{\partial}{\phi}_{1}, 
\ .., {\partial}s/{\partial}{\phi}_{\nu})\bigr)$
is then the conjugate, ${\theta} \ (=({\theta}_{1}, \ .,{\theta}_{\nu}))$ of ${\phi}$  
and, by Eqs. (2.2), (2.3), (2.7) and  (2.8), it is given explicitly by the formula
$${\theta}=s^{\prime}({\phi})=
{\beta}(1,-{\mu}+u^{2}/2,-u_{1},.  .,-u_{d})\eqno(2.9)$$
or, more compactly,
$${\theta}=s^{\prime}({\phi})={\beta}(1,-{\mu}+u^{2}/2,-u).\eqno(2.9a)$$
\vskip 0.2cm
The thermodynamic conjugate, ${\pi}$, of $s$ is the function of ${\theta}$ defined 
by the formula
$${\pi}({\theta})={\rm sup}_{\phi}\bigl(s({\phi})-{\theta}.{\phi}\bigr).\eqno(2.10)$$
We assume that all parts of this system are in a single thermodynamical phase at all 
times and consequently that the function $s^{\prime}$ is invertible and the supremum 
on the r.h.s. of Eq. (2.10) is attained when ${\phi}=[s^{\prime}]^{-1}({\theta})$. 
Hence
$${\pi}({\theta})=s({\phi})-{\theta}.{\phi}, {\rm with} \  
s^{\prime}({\phi})={\theta},\eqno(2.11)$$
which, together with Eqs. (2.2)-(2.4), (2.8) and (2.9a), implies that 
$${\pi}({\theta})={\beta}p.\eqno(2.12)$$
Furthermore, it follows from Eq. (2.11) that 
$${\pi}^{\prime}({\theta})=-{\phi}.\eqno(2.13)$$
Consequently, the Hessians $s^{{\prime}{\prime}}({\phi}) \ 
\bigl(=\bigl([{\partial}^{2}s/{\partial}{\phi}_{j}{\partial}{\phi}_{k}]\bigr)\bigr)$ 
and ${\pi}^{{\prime}{\prime}}({\theta})$ are related by the equations 
$${\pi}^{{\prime}{\prime}}({\theta})s^{{\prime}{\prime}}({\phi})
=s^{{\prime}{\prime}}({\phi}){\pi}^{{\prime}{\prime}}({\theta})=-I,$$
i.e.
$${\pi}^{{\prime}{\prime}}({\theta})=-s^{{\prime}{\prime}}({\phi})^{-1}.
\eqno(2.14)$$
\vskip 0.5cm
\centerline {\bf 2.2. The Navier-Stokes Equations.}
\vskip 0.3cm
These hydrodynamical equations comprise local conservation laws for the energy, 
mass and momentum, together with constitutive equations for the associated fluxes. 
The local conservation laws are
$${{\partial}e\over {\partial}t}+{\nabla}.q=0; \ 
{{\partial}{\rho}\over {\partial}t}+{\nabla}.j=0; \ {\rm and} \ 
{{\partial}j\over {\partial}t}+{\nabla}.{\tau}=0,\eqno(2.15)$$
where $q$ is the energy current, ${\tau}$ is the stress tensor $[{\tau}_{kl}]$ and 
$({\nabla}.{\tau})_{k}:={\partial}{\tau}_{kl}/{\partial}x_{l}$. The constitutive 
equations are then [3]
$$q=\bigl({\varepsilon}+{1\over 2}{\rho}u^{2}\bigr)u-
{\sigma}.u+{\kappa}{\nabla}{\beta},\eqno(2.16)$$
$${\tau}=pI+{\rho}uu-{\sigma}\eqno(2.17)$$
and
$${\sigma}={\gamma}_{1}\bigl(Du-2d^{-1}({\nabla}.u)I\bigr)+{\gamma}_{2}
({\nabla}.u)I,\eqno(2.18)$$ 
where
$$(Du)_{kl}={{\partial}u_{k}\over {\partial}x_{l}}+
{{\partial}u_{l}\over {\partial}x_{k}},\eqno(2.19)$$
$uu$ is the dyadic whose $kl$ component is $u_{k}u_{l}$ and ${\kappa}, \ 
{\gamma}_{1}$ and ${\gamma}_{2}$ are positive, scalar valued functions of 
${\beta}$ and ${\mu}$ that represent the thermal conductivity, the bulk viscosity and 
the shear viscosity, respectively.
\vskip 0.2cm
Thus, in view of the one-to-one correspondence between the variables $(e,{\rho},j)$ 
and $({\beta},{\mu},u)$, the combination of the conservation laws (2.15) and the 
constitutive equations (2.16)-(2.18) describe an autonomous evolution of the 
hydrodynamical variables. Further, as remarked at the beginning of Section 2, the 
boundary conditions are determined by the reservoirs with which the system is in 
contact. Specifically, we assume that the contact is restricted to the surface, 
${\partial}{\Omega}$, of ${\Omega}$ and the boundary conditions are taken to be 
the following ones.
\vskip 0.2cm\noindent
(a) $u=0$ on ${\partial}{\Omega}$. 
\vskip 0.2cm\noindent
(b) At any point of ${\partial}{\Omega}$ that is in contact with a reservoir, the values 
of the temperature and chemical potential of ${\Sigma}$ are just those of the 
reservoir.
\vskip 0.2cm\noindent 
(c) Any part of ${\partial}{\Omega}$ not in contact with a reservoir is insulated and 
the normal part of ${\nabla}{\theta}$ is zero there. 
\vskip 0.2cm
We now compactify the hydrodynamical equations  by expressing them in terms of 
the triples ${\phi}:=(e,{\rho},j)$ and ${\chi}:=(q,j,{\tau})$, the latter of which we 
shall term the {\it flux}. Inserting subscripts $t$ to indicate time dependence of the 
hydrodynamical variables, the local conservation laws (2.15) may be expressed as the 
single equation
$${{\partial}{\phi}_{t}\over {\partial}t}+{\nabla}.{\chi}_{t}=0,\eqno(2.20)$$
and Eqs. (2.16)-(2.18) comprise a constitutive equation for ${\chi}_{t}$ of the form
$${\chi}_{t}(x)={\cal G}({\phi}_{t}:x).\eqno(2.21)$$
By Eqs. (2.20) and (2.21), ${\phi}_{t}$ evolves according to the autonomous 
equation of motion 
$${{\partial}{\phi}_{t}(x)\over {\partial}t} ={\cal F}({\phi}_{t};x),\eqno(2.22)$$
where
$${\cal F}({\phi}_{t};x):=-{\nabla}.{\cal G}({\phi}_{t};x).\eqno(2.23)$$
We assume that the equation of motion (2.22) has a unique stationary solution, subject 
to the prevailing boundary conditions, described above. We denote this solution by 
${\overline {\phi}}(x)=({\overline e}(x),{\overline {\rho}}(x),{\overline j}(x)$ and 
denote the corresponding steady state value of ${\theta}(x)$ by 
${\overline {\theta}}(x)$.  
\vskip 0.3cm
{\it The Perturbed flow.} We assume that the stationary flow is stable under 
\lq small\rq\ perturbations 
${\delta}{\phi}_{t}(x)=\bigl({\delta}e_{t}(x),{\delta}{\rho}_{t}(x),
{\delta}j_{t}(x)\bigr)$ of ${\overline {\phi}}(x)$, that preserve the boundary 
conditions. The resultant linearised equation of motion is then
$${{\partial}\over {\partial}t}{\delta}{\phi}_{t}(x)=
({\cal L}{\delta}{\phi}_{t})(x)
:={{\partial}\over {\partial}{\lambda}}
{\cal F}\bigl({\overline{\phi}}+
{\lambda}{\delta}{\phi}_{t};x\bigr)_{{\vert}{\lambda}=0}.\eqno(2.24)$$
Thus, by Eqs. (2.23) and (2.24),
$${\cal L}{\delta}{\phi}_{t}=
-{\nabla}.{\cal K}{\delta}{\phi}_{t},\eqno(2.25)$$
where
$$({\cal K}{\delta}{\phi}_{t})(x)={{\partial}\over {\partial}{\lambda}}
{\cal G}\bigl({\overline{\phi}}+
{\lambda}{\delta}{\phi}_{t};x\bigr)_{{\vert}{{\lambda}=0}}.\eqno(2.26)$$
Further, by Eqs. (2.21) and (2.26), the increment ${\delta}{\chi}_{t}$ in ${\chi}_{t}$ 
is given by the formula
$${\delta}{\chi}_{t}(x)={\cal K}{\delta}{\phi}_{t}(x).\eqno(2.27)$$
\vskip 0.2cm
In anticipation of the demands of the macrostatistical picture of the system, 
formulated in the next Section, we assume that ${\delta}{\phi}_{t}$ is a distribution, 
in the sense of L. Schwartz [6]. Thus, since ${\delta}{\phi}_{t}=
({\delta}e_{t},{\delta}{\rho}_{t},{\delta}j_{t})$, where the first two components are 
scalar fields and the third is a vector field, we assume that ${\delta}{\phi}_{t}$ is an 
element of the space ${\tilde {\cal D}}^{\prime}({\Omega})$, defined at the end of 
Section 1. Thus, employing the notations introduced there, it follows from the 
definition of ${\delta}{\phi}_{t}$ that, for $F=(f,g,h)$, with $f,g{\in}{\cal 
D}({\Omega})$ and $h{\in}{\cal D}_{V}({\Omega})$,
$${\delta}{\phi}_{t}(F)={\delta}e_{t}(f)+{\delta}{\rho}_{t}(g)+{\delta}j_{t}(h),
\eqno(2.28)$$
where the three terms on the r.h.s. are the integrals of ${\delta}e_{t}, \ 
{\delta}{\rho}_{t}$ and ${\delta}j_{t}$ against $f,g$ and $h$, respectively.    
\vskip 0.2cm
We now assume that the linear operator ${\cal L}$, defined in Eq. (2.24), is the 
generator of a one parameter semigroup 
$T:={\lbrace}T_{t}{\vert}t{\in}{\bf R}_{+}{\rbrace}$  of linear transformations of 
${\tilde {\cal D}}^{\prime}({\Omega})$. Hence, by Eq. (2.24),
$${\delta}{\phi}_{t}=T_{t}{\delta}{\phi}_{0} \ {\forall}\ t{\in}
{\bf R}_{+},\eqno(2.29)$$ 
from which it follows that
$${\delta}{\phi}_{t}(F)={\delta}{\phi}\bigl(T^{\star}_{t}F) \ {\forall} \ 
F{\in}{\tilde {\cal D}}({\Omega}). \ t{\in}{\bf R}_{+},\eqno(2.30)$$
where $T^{\star}:={\lbrace}T_{t}^{\star}{\vert}t{\in}{\bf R}_{+}{\rbrace}$ is the 
dual of the semigroup $T$: its generator, ${\cal L}^{\star}$, is the dual of ${\cal L}$. 
We assume the dissipativity condition that, for all perturbations ${\delta}{\phi}, \ 
T_{t}{\delta}{\phi}$ tends to zero, in the 
${\tilde {\cal D}}^{\prime}({\Omega})$ topology, as $t{\rightarrow}{\infty}$. As 
the space ${\tilde {\cal D}}({\Omega})$ is reflexive, this is equivalent to the 
assumption that
$${\tilde {\cal D}}({\Omega}):{\rm lim}_{t\to\infty}T_{t}^{\star}F=0 \ {\forall} \ 
F{\in}{\tilde {\cal D}}({\Omega}).\eqno(2.31)$$
\vskip 0.3cm
{\it Re-expression of NS in terms of ${\theta}$.} Since the constitutive equations 
(2.16)-(2.19) relate the heat and mass currents, as well as the stress tensor, directly to 
${\theta}$, it is sometimes convenient to express the r.h.s.\rq s of Eqs. (2.22)-(2.24) in 
terms of this variable rather than ${\phi}$. Thus we rewrite Eq. (2.22) as
$${{\partial}{\phi}_{t}\over {\partial}t}=-{\nabla}.{\Psi}({\theta}_{t};x),
\eqno(2.32)$$
where ${\Psi}({\theta}_{t};x)={\cal F}({\phi}_{t};x)$. Eq. (2.24) for the perturbed 
flow then becomes
$${{\partial}\over {\partial}t}{\delta}{\phi}_{t}={\Lambda}{\delta}{\theta}_{t}:=
-{{\partial}\over {\partial}{\lambda}}{\nabla}.{\Psi}({\overline {\theta}}+{\lambda}
{\delta}{\theta}_{t};x)_{{\vert}{\lambda}=0}.\eqno(2.33)$$ 
Further, by Eq. (2.13) and the equivalence of Eqs. (2.24) and (2.33),
$${\Lambda}{\delta}{\theta}={\cal L}{\delta}{\phi}=-{\cal L}
{\pi}^{{\prime}{\prime}}({\overline {\theta}}){\delta}{\theta}$$
for all perturbations ${\delta}{\theta}$ of ${\overline {\theta}}$ and hence 
$${\Lambda}=-{\cal L}{\pi}^{{\prime}{\prime}}({\overline {\theta}}).
\eqno(2.34)$$
It follows immediately from this equation and the symmetry of 
${\pi}({\theta})^{{\prime}{\prime}}$ that the dual, ${\Lambda}^{\star}$, of 
${\Lambda}$ is related to that, ${\cal L}^{\star}$, of ${\cal L}$ by the formula
$${\Lambda}^{\star}=-{\pi}^{{\prime}{\prime}}({\overline {\theta}})
{\cal L}^{\star}.\eqno(2.35)$$
In particular, as we shall show in Appendix A, it follows from our definitions that the 
equilibrium form, ${\Lambda}_{\rm eq}^{\star}$, of ${\Lambda}^{\star}$ is given 
by the 
equation
$${\Lambda}_{\rm eq}^{\star}(f,g,h)=$$
$$-\Bigl({\kappa}{\Delta}f-{\beta}^{-1}{\varepsilon}{\nabla}.h, \ 
-{\beta}^{-1}{\rho}{\nabla}.h, \ $$
$$-{\beta}^{-1}{\varepsilon}{\nabla}f-{\beta}^{-1}{\rho}
{\nabla}g+{\beta}^{-1}{\gamma}_{1}{\nabla}.(Dh-2d^{-1}({\nabla}.h)I)
+{\beta}^{-1}{\gamma}_{2}{\nabla}({\nabla}.h)\Bigr)$$
$${\forall} \ (f,g,h){\in}{\tilde {\cal D}}){\Omega}).
\eqno(2.36)$$
 \vskip 0.5cm
\centerline {\bf 3. The Stochastic Fluctuation Process} 
\vskip 0.3cm
According to elementary statistical mechanics, the thermodynamic and hydrodynamic 
variables undergo fluctuations, which are not taken into account in the 
phenomenological picture of the previous Section. 
We now seek to provide a general description of the hydrodynamical fluctuations 
about nonequilibrium steady states by treating the fields ${\phi}=(e,{\rho},j)$ and 
${\chi}=(q,j,{\tau})$ as expectation values of random fields ${\hat {\phi}}=
({\hat e},{\hat {\rho}},{\hat j})$ and ${\hat {\chi}}=({\hat q},{\hat j},{\hat {\tau}})$,
respectively, where the tensor ${\hat {\tau}}$, like the phenomenological ${\tau}$, is 
symmetric\footnote*{This symmetry property may be regarded vas basic, as it 
prevails in standard micrroscopic pictures of stress tensors.}. We assume that these 
fields satisfy the local conservation law given by the canonical analogue of Eq. (2.20), 
namely
$${{\partial}{\hat {\phi}}_{t}\over {\partial}t}+{\nabla}.{\hat 
{\chi}}_{t}=0.\eqno(3.1)$$
The differences between the random fields ${\hat {\phi}}_{t}$ and ${\hat 
{\chi}}_{t}$ and their classical expectation values then represent the hydrodynamical 
fluctuations. In a standard way, we normalise them by a factor ${\cal N}^{1/2}$, 
where ${\cal N}$ is chosen to be the ratio of characteristic values of corresponding 
macroscopic and microscopic quantities. In the present situation, where we are 
formulating the model on a macroscopic scale, a natural choice for ${\cal N}$ is the 
reciprocal, $k_{B}^{-1}$, of Boltzmann\rq s constant, which arises in Einstein\rq s 
formula, $P={\rm const.}{\rm exp}(S/k_{B})$, relating the equilibrium probability 
distribution of the macroscopic variables to the entropy $S$. Thus, we define the 
fields representing the fluctuations of ${\phi}_{t}$ and ${\chi}_{t}$ about their 
steady state values to be  
$${\xi}_{t}(x)=k_{B}^{-1/2}\bigl({\hat {\phi}}_{t}(x)-
{\overline {\phi}}\bigr)\eqno(3.2)$$
and 
$${\eta}_{t}(x)=k_{B}^{1/2}\bigl({\hat {\chi}}_{t}(x)-
{\overline {\chi}}\bigr),\eqno(3.3)$$
respectively. Hence, defining
$${\zeta}_{t,s}:=\int_{s}^{t}du{\eta}_{u},\eqno(3.4)$$
it follows from Eqs. (3.1)-(3.4) that
$${\xi}_{t}-{\xi}_{s}=-{\nabla}.{\zeta}_{t,s} \ {\forall} \ t,s{\in}{\bf R}.
\eqno(3.5)$$
\vskip 0.2cm
In accordance with the general requirements of field theories [8], we assume that 
these fields are distributions, in the sense of L. Schwartz [6]. Specifically, in the 
notation of the last Note of Section 1, we assume that 
${\xi}_{t}{\in}{\tilde {\cal D}}^{\prime}({\Omega})$ and ${\eta}_{t}{\in} {\check 
{\cal D}}^{\prime}({\Omega})$. We denote the smeared fields obtained by 
integrating ${\xi}_{t}$ and ${\zeta}_{t,s}$ against test functions $F \ 
\bigl({\in}{\tilde {\cal D}}({\Omega})\bigr)$ and $G \ \bigl({\in}{\check {\cal 
D}}({\Omega})\bigr)$ by ${\xi}_{t}(F)$ and ${\eta}_{t}(G)$, respectively; 
\vskip 0.2cm
We aim to derive the stochastic process executed by the fluctuation fields ${\xi}$ and 
${\zeta}$ from the assumptions of 
\vskip 0.2cm\noindent 
(a) a generalised version of Onsager\rq s regression hypothesis,
\vskip 0.2cm\noindent
(b) a macrostatistical local equilibrium hypothesis and
\vskip 0.2cm\noindent
(c) a chaoticty hypothesis for the random currents and stresses. 
\vskip 0.2cm\noindent
Further, as the continuum model harbours only macroscopic variables, we take any 
counterparts to microscopic correlation lengths and memory times of this model to be 
zero. 
\vskip 0.5cm
\centerline {\bf 3.1. Regression Hypothesis.} 
\vskip 0.3cm
The regression hypothesis signifies that  the fluctuations of the hydrodynamical 
observables evolve, from a given starting point, according to the same dynamical law 
that governs the evolution of small perturbations of those variables from their steady 
state values. Thus, noting the formula (2.30) for the latter evolution and denoting 
conditional expectations of the stochastic variables with respect to the value of the 
field ${\xi}$ at time $t_{0}$ by 
$E(.{\vert}{\xi }_{t_{0}})$, we assume that
$$E({\xi}_{t}(F){\vert}{\xi }_{t_{0}})= {\xi}_{t_{0}}
\bigl((T^{\star}(t-t_{0})F\bigr) \ {\forall} \ 
F{\in}{\tilde {\cal D}}({\Omega}), \ t_{0},t({\geq}t_{0}){\in}{\bf R}.\eqno(3.6)$$
Since the process ${\xi}_{t}$ is stationary, an immediate consequence of this formula 
is that
$$E\bigl({\xi}_{t}(F){\xi}_{t^{\prime}}(F^{\prime})\bigr)=
E\bigl({\xi}[T^{\star}(t-t^{\prime})F]{\xi}(F^{\prime})\bigr) \ {\forall} \ 
F,F^{\prime}{\in} 
{\tilde {\cal D}}({\Omega}), \ t,t^{\prime}({\leq}t){\in}{\bf R}.\eqno(3.7)$$
\vskip 0.2cm
We now note that, by Eq. (2.27), the increment in the integrated phenomenological 
flux, $\int_{s}^{t}du{\chi}_{u}$, due to a perturbation ${\delta}{\phi}$ of ${\phi}$ 
is $\int_{s}^{t}du{\cal K}{\delta}{\phi}_{u}$. Correspondingly, as the 
phenomenological dynamics of the model is secular, we designate the secular part of 
the integrated fluctuation flux, ${\zeta}_{t,s}$, to be 
$${\zeta}_{t,s}^{\rm sec}:=\int_{s}^{t}du{\cal K}{\xi}_{u}.\eqno(3.8)$$ 
We then define the remaining part, ${\tilde {\zeta}}_{t,s}$, of ${\zeta}_{t,s}$ to be 
the stochastic part of the integrated flux. Thus
$${\tilde {\zeta}}_{t,s}:={\zeta}_{t,s}-\int_{s}^{t}du{\cal K}{\xi}_{u}.\eqno(3.9)$$
We shall presently show, in Prop. 3.1, that this field does indeed enjoy strong, 
Wiener-like, stochastic properties.
\vskip 0.2cm
By Eqs. (2.25), (3.5) and (3.9),
$${\xi}_{t}(F)-{\xi}_{s}(F)=\int_{s}^{t}du{\xi}_{u}({\cal L}^{\star}F)+w_{t,s}(F) 
\ {\forall} \ s,t{\in}{\bf R}, \ F{\in}{\tilde {\cal D}}({\Omega}),\eqno(3.10)$$
i.e.
$${\xi}_{t}-{\xi}_{s}=\int_{s}^{t}du{\cal L}{\xi}_{u}+w_{t,s},\eqno(3.10a)$$
where
$$w_{t,s}(F)={\tilde {\zeta}}_{t,s}({\nabla}F).\eqno(3.11)$$
The following Proposition, which was proved in [5, Prop. 5.1], establishes that $w$ 
simulates a Wiener process, at least as far as its two-point function is concerned; and 
hence that Eq. (3.10a) is an integrated Langevin equation.
\vskip 0.3cm
{\bf Proposition 3.1.} {\it Under the assumption of the regression hypothesis, $w$ has 
the following properties.
\vskip 0.2cm\noindent
(i) $$E\bigl(w_{t,s}(F){\xi}_{u}(F^{\prime})\bigr)=0 \ {\forall} \ t{\geq}s{\geq}u, \ 
F,F^{\prime}{\in}{\tilde {\cal D}}({\Omega}).\eqno(3.12)$$
\vskip 0.2cm\noindent 
(ii) $$E\bigl(w_{t,s}(F)w_{t^{\prime},s^{\prime}}(F^{\prime})\bigr)=
E\bigl({\xi}({\cal L}^{\star}F){\xi}(F^{\prime})+
{\xi}(F){\xi}({\cal L}^{\star}F^{\prime})\bigr)
{\vert}[s,t]{\cap}[s^{\prime},t^{\prime}]{\vert}$$
$${\forall} \ s,t({\geq}s), \ s^{\prime},t^{\prime}({\geq}s) \ {\in}{\bf R}, \ 
F,F^{\prime}{\in}{\tilde {\cal D}}({\Omega}),\eqno(3.13)$$
where ${\vert}I{\vert}$ denotes the length of an interval $I$ in ${\bf R}$.}
\vskip 0.5cm
\centerline {\bf  3.2. The Chaoticity Hypothesis.} 
\vskip 0.3cm
We now strengthen Prop. 3.1 by the assumption that the stochastic part, ${\tilde 
{\zeta}}$ of the integrated fluctuation flux is chaotic in the sense that its space-time 
correlations are of microscopic range, idealised here as zero range on our macroscopic 
scale. This assumption is designed to represent Boltzmann\rq s molecular chaos 
hypothesis, as transferred to the stochastic flux. Here we take it to signify that the 
field ${\tilde {\zeta}}$ is Gaussian, due to statistical independence of its values in 
disjoint space-time regions. Thus, our chaoticity hypothesis comprises the following 
conditions.
\vskip 0.2cm\noindent
(C.1) The process ${\tilde {\zeta}}$ is Gaussian, and
\vskip 0.2cm\noindent
(C.2) $E\bigl({\tilde {\zeta}}_{s,t}(G)
{\tilde {\zeta}}_{s^{\prime},t^{\prime}}(G^{\prime})\bigr)$ vanishes if the 
intersection of either the intervals $[s,t]$ and $[s^{\prime},t^{\prime}]$ or of the 
supports of $G$ and $G^{\prime}$  is empty. 
\vskip 0.2cm 
Further, we supplement these conditions by the following continuity assumption.
\vskip 0.2cm\noindent
${\cal C}$. The correlation function of (C.2) is continuous in its time variables.
\vskip 0.2cm
The following Proposition was proved in [5, Prop. 5.2].
\vskip 0.3cm
{\bf Proposition 3.2.} {\it Under the assumption of (C.2) and ${\cal C}$, the 
two=point function of ${\tilde {\zeta}}$ takes the following form. 
$$E\bigl({\tilde {\zeta}}_{t,s}(G)
{\tilde {\zeta}}_{s^{\prime},t^{\prime}}(G^{\prime})\bigr)=
{\Gamma}(G,G^{\prime}){\vert}[s,t]{\cap}[s^{\prime},t^{\prime}]{\vert} 
\ {\forall} \ G,G^{\prime}{\in}{\tilde {\cal D}}^{\prime}({\Omega}), \ 
s,t,s^{\prime},t^{\prime}{\in}{\bf R},\eqno(3.14)$$
where ${\Gamma}$ is a continuous bilinear form on 
${\check {\cal D}}({\Omega}){\otimes}{\check {\cal D}}({\Omega})$, whose 
support lies in the region 
${\lbrace}(x,x^{\prime}){\in}{\Omega}^{2}{\vert}x=x^{\prime}{\rbrace}$.}
 \vskip 0.3cm
We shall derive the explicit form of ${\Gamma}$ from the local equilibrium 
hypothesis in Section 3.3. The following Corollary to Prop. 3.2 is a simple 
consequence of Eq. (3.13), assumption (C.1) and Prop. (3.2).
\vskip 0.3cm
{\bf Corollary 3.3.} {\it Under the above assumptions, the process $w$ is Gaussian 
with zero mean and its two-point function takes the form
$$E\bigl(w_{t,s}(F)w_{t^{\prime},s^{\prime}}(F^{\prime})\bigr)=
{\Gamma}({\nabla}F,{\nabla}F^{\prime}) 
{\vert}[s,t]{\cap}[s^{\prime},t^{\prime}]{\vert} \ {\forall} \ F,F^{\prime}{\in}
{\tilde {\cal D}}({\Omega}), \ t,s,t^{\prime},s^{\prime}{\in}{\bf R}.\eqno(3.15)$$}
\vskip 0.3cm
The following Proposition was shown in [5, Prop. 5.5] to ensue from the Langevin 
equation (3.10a),  Prop. 3.1 (i) and Cor. 3.3.
\vskip 0.3cm
{\bf Proposition 3.4.} {\it Under the above assumptions, ${\xi}$ is a Gaussian 
Markov process, and the fields $w_{t,s}$ and ${\xi}_{u}$ are statistically 
independent of one another if $s$ and $t$ are greater than or equal to $u$.}
\vskip 0.5cm
\centerline {\bf 3.3 Equilibrium and Local Equilibrium Conditions.}
\vskip 0.3cm
{\it Equilibrium Statistics of ${\xi}$.} We base our formulation of these statistics on 
the canonical version of Einstein\rq s formula for the probability distribution, $P$ of 
the hydrodynamical variables. This is given formally by the equation
$$P_{eq}={\rm const.}{\rm exp}\bigl(k_{B}^{-1}
\int_{\Omega}dx[s({\hat {\phi}}(x)-{\overline {\theta}}.{\hat {\phi}}(x)]\bigr),$$
i.e., by Eq. (3,2),
$$P_{eq}={\rm const.}{\rm exp}\bigl(k_{B}^{-1}\int_{\Omega}dx
[s({\overline {\phi}}+k_{B}^{1/2}{\xi}(x))-
{\overline {\theta}}.({\overline {\phi}}+k_{B}^{1/2}{\xi}(x)]\bigr).\eqno(3.16)$$
Since the integrand of this formula is maximised at ${\xi}=0$, its ratio to $k_{B}$ 
reduces to $\bigl({\xi}(x).s^{{\prime}{\prime}}({\overline 
{\theta}}){\xi}(x)\bigr)/2$ in the limit 
$k_{B}{\rightarrow}0$, which we term the {\it Botzmann} limit: here the dot denotes 
the ${\bf R}^{\nu}$ scalar product. Thus, operating henceforth in this limit, the 
equilibrium characteristic function for the fluctuation field ${\xi}$ is given formally 
by  
$$E_{\rm eq}\bigl({\rm exp}[i{\xi}(F)]\bigr)=
{\rm const.}{{\int}{\cal D}{\xi}(x){\rm exp}
\bigl[\int_{\Omega}dx\bigl({\xi}(x).s^{{\prime}{\prime}}
({\overline {\phi}}){\xi}(x)/2+i{\xi}(x).F(x)\bigr)\bigr]
\over {\int}{\cal D}{\xi}(x){\rm exp}
\bigl[\int_{\Omega}dx\bigl({\xi}(x).s^{{\prime}{\prime}}
({\overline {\phi}}){\xi}(x)\bigr)/2\bigr)\bigr]} \ {\forall} \ 
F{\in}{\tilde {\cal D}}({\Omega}),\eqno(3.17)$$
where ${\cal D}{\xi}$ denotes functional integration w.r.t. the field ${\xi}$. This 
formula may properly be defined by resolving ${\Omega}$ into a set of cells, 
${\Delta}J$, denoting the values of ${\xi}$ and $F$ at the centre of ${\Delta}J$ by 
${\xi}_{J}$ and $F_{J}$, respectively, and then expressing Eq. (3.17) as 
$$ E_{\rm eq}\bigl({\rm exp}[i{\xi}(F)]\bigr)=
{\rm lim}_{\Delta}{\Pi}_{J}\Bigl[{\int_{{\bf R}^{\nu}}d{\xi}_{J}{\exp}
\bigl({\xi}_{J}.s^{{\prime}{\prime}}
({\overline {\theta}}){\xi}_{J}/2+i{\xi}_{J}.F_{J}\bigr)\over 
\int_{{\bf R}^{\nu}}d{\xi}_{J}{\exp}\bigl({\xi}_{J}.s^{{\prime}{\prime}}
({\overline {\theta}}){\xi}_{J}/2\bigr)}\Bigr],\eqno(3.17a)$$
where ${\rm lim}_{\Delta}$ is the limit in which the cells shrink to points. It now 
follows easily from Eqs.(2.14) and (3.17a) that, in the notation specified at the end of 
Section 1,
$$E_{\rm eq}\bigl({\rm exp}[i{\xi}(F)]\bigr)={\exp}\bigl(-{1\over 2}
{\langle}F,{\pi}^{{\prime}{\prime}}({\overline {\theta}})F{\rangle}\bigr) \ {\forall} 
\ F{\in}{\tilde {\cal D}}({\Omega}).\eqno(3.18)$$ 
Thus, at equilibrium, ${\xi}$ is a Gaussian random field, with zero mean and two-
point function given by the formula
$$E_{\rm eq}\bigl({\xi}(F))^{2}\bigr)=
{1\over 2}{\langle}F,{\pi}^{{\prime}{\prime}}
({\overline {\theta}})F{\rangle} \ {\forall} \ F{\in}{\tilde {\cal D}}({\Omega}),$$
or equivalently, by polarisation,
$$E_{\rm eq}{\langle}{\xi}(F){\xi}(F^{\prime})\bigr)=
{\langle}F,{\pi}^{{\prime}{\prime}}({\overline {\theta}})F^{\prime}){\rangle} \  
{\forall} \ F,F^{\prime}{\in}{\tilde {\cal D}}({\Omega}).\eqno(3.19)$$
\vskip 0.3cm
{\it Equilibrium two-point function for ${\tilde {\zeta}}$.} By Eqs. (2.34), (3.11), 
(3.13) and (3.19),
$$E_{\rm eq}\bigl({\tilde {\zeta}}_{t,s}({\nabla}F)
{\tilde {\zeta}}_{t^{\prime},s^{\prime}}(F^{\prime})\bigr)=
\bigl[{\langle}{\Lambda}_{\rm eq}^{\star}F,F^{\prime}{\rangle}+
{\langle}F,{\Lambda}_{\rm eq}^{\star}F^{\prime}{\rangle}\bigr]
{\vert}[s,t]{\cap}[s^{\prime},t^{\prime}]{\vert}$$
$${\forall} \ F,F^{\prime}{\in}
{\tilde {\cal D}}({\Omega}), \  s,t,s^{\prime},t^{\prime}{\in}{\bf R}.\eqno(3.20)$$
Hence, for $F=(f,g,h)$ and $F^{\prime}=(f^{\prime},g^{\prime},h^{\prime})$, it 
follows from Eqs. (2.36) and (3.20) that
$$E_{\rm eq}\bigl({\tilde {\zeta}}_{t,s}({\nabla}F)
{\tilde {\zeta}}_{t^{\prime},s^{\prime}}({\nabla}F^{\prime})\bigr)=$$
$$\bigl[2{\kappa}{\langle}{\nabla}f,{\nabla}f^{\prime}{\rangle}+
{\beta}^{-1}{\gamma}_{1}\bigl({\langle}Dh,Dh{\rangle}-
2d^{-1}{\langle}{\nabla}.h,{\nabla}.h^{\prime}{\rangle}\bigr)
+2{\beta}^{-1}{\gamma}_{2}{\langle}
{\nabla}.h,{\nabla}.h^{\prime}{\rangle}\bigr]{\times}$$
$${\vert}[s,t]{\cap}[s^{\prime},t^{\prime}]{\vert}.\eqno(3.21)$$
\vskip 0.2cm
We now express ${\tilde {\zeta}}_{t,s}$ in the form 
$${\tilde {\zeta}}_{t,s}=
({\tilde {\zeta}}_{t,s}^{(1)},{\tilde {\zeta}}_{t,s}^{(2)},
{\tilde {\zeta}}_{t,s}^{(3)}),\eqno(3.22)$$ 
where the components are the time integrals of the stochastic fluctuations of 
$q,j,{\tau}$ and therefore lie in ${\cal D}_{V}^{\prime}({\Omega}), \ 
{\cal D}_{V}^{\prime}({\Omega})$ and  ${\cal D}_{T}^{\prime}({\Omega})$, 
respectively. Thus, 
$${\tilde {\zeta}}_{t,s}({\nabla}F)={\tilde {\zeta}}_{t,s}^{(1)}({\nabla}f)+
{\tilde {\zeta}}_{t,s}^{(2)}({\nabla}g)+{\tilde {\zeta}}_{t,s}^{(3)}({\nabla}h)$$
and similarly
$${\tilde {\zeta}}_{t,s}({\nabla}F^{\prime})=
{\tilde {\zeta}}_{t,s}^{(1)}({\nabla}f^{\prime})+
{\tilde {\zeta}}_{t,s}^{(2)}({\nabla}g^{\prime})+
{\tilde {\zeta}}_{t,s}^{(3)}({\nabla}h^{\prime}).$$ 
The substitution of these last two equations in Eq. (3.21) yields the formulae 
$$E_{\rm eq}\bigl({\tilde {\zeta}}_{t,s}^{(1)}({\nabla}f)
{\tilde {\zeta}}_{t^{\prime},s^{\prime}}^{(1)}({\nabla}f^{\prime})\bigr)=
2{\kappa}{\langle}{\nabla}f,{\nabla}f^{\prime}{\rangle}{\vert}[s,t]{\cap}
[s^{\prime},t^{\prime}]{\vert},\eqno(3.23)$$ 
and
$$E_{\rm eq}\bigl({\tilde {\zeta}}_{t,s}^{(3)}({\nabla}h)
{\tilde {\zeta}}_{t,s}^{(3)}({\nabla}h^{\prime})\bigr)=
\bigl[2{\beta}^{-1}{\gamma}_{1}\bigl({\langle}Dh,Dh^{\prime}{\rangle}
-2d^{-1}{\langle}{\nabla}.h,{\nabla}.h^{\prime}{\rangle}\bigr)+
{\beta}^{-1}{\gamma}_{2}{\langle}{\nabla}.h,{\nabla}.h^{\prime})
{\rangle}\bigr]$$
$${\vert}[s,t]{\cap}[s^{\prime},t^{\prime}]{\vert}.\eqno(3.24)$$
and implies that all other two-point functions for the components of ${\tilde 
{\zeta}}_{t,s}({\nabla}F)$ are zero 
\vskip 0.2cm
The following lemma for the generalised functions ${\tilde {\zeta}}_{t,s}^{(1)}$ and 
${\tilde {\zeta}}_{t,s}^{(3)}$ on ${\Omega}$ corresponding to the distributions 
denoted by the same symbols will proved in Appendix B. 
\vskip 0.3cm
{\bf Lemma 3.5.} {\it Under the above  assumptions, supplemented by the condition 
that the equilibrium two-point function of ${\tilde {\zeta}}$ is locally translationally 
and rotationally invariant, and indicating the components of $X$-vectors by 
subscripts $i,j,l,m$ 
$$E_{\rm eq}\bigl({\tilde {\zeta}}_{t,s;i}^{(1)}(x)
{\tilde {\zeta}}_{t^{\prime},s^{\prime};j}^{(1)}
(x^{\prime})\bigr)=2{\kappa}{\delta}(x-x^{\prime})
{\vert}[s,t]{\cap}[s^{\prime},t^{\prime}{\vert},\eqno(3.25)$$
$$E_{\rm eq}\bigl({\tilde {\zeta}}_{t,s;il}^{(3)}(x)
{\tilde {\zeta}}_{t^{\prime},s^{\prime};jm}^{(3)}(x^{\prime})\bigr)=$$
$$2{\beta}^{-1}\bigl({\gamma}_{1}({\delta}_{ij}{\delta}_{lm}+{\delta}_{im}
{\delta}_{jm}-2d^{-1}{\delta}_{il}{\delta}_{jm}
{\delta}_{jl})+{\gamma}_{2}
{\delta}_{il}{\delta}_{jm}\bigr){\delta}(x-x^{\prime})
{\vert}[s,t]{\cap}[s^{\prime},t^{\prime}]{\vert}\eqno(3.26)$$
and all other two-point functions of the components of ${\tilde {\zeta}}_{t,s}$ are 
zero.This result concurs with that of Landau [3, Eqs. (132.11-13)].}
\vskip 0.3cm
In order to re-express the two-point functions of ${\tilde {\zeta}}$ in terms of 
smeared fields, we denote elements $G$ and $G^{\prime}$ of 
${\check{\cal D}}({\Omega})$ by triples $(a,b,c)$ and 
$(a^{\prime},b^{\prime},c^{\prime})$, respectively, where the first two components 
of both $G$ and $G^{\prime}$ lie in ${\cal D}_{V}({\Omega})$ and the third lie in 
${\cal D}_{T}({\Omega})$. We then define the elements $c^{(1)}, \  
c^{{\prime}(1)}$ of ${\cal D}_{T}({\Omega})$ and $c^{(2)}, \  c^{{\prime}(2)}$ 
of ${\cal D}({\Omega})$ by the formulae
$$c_{ij}^{(1)}:=c_{ij}+c_{ji}; \ 
c_{ij}^{{\prime}(1)}:=c_{ij}^{\prime}+c_{ji}^{\prime}\eqno(3.27)$$
and
$$c^{(2)}:=c_{jj}; \ c^{{\prime}(2)}=c_{jj}^{\prime}.\eqno(3.28)$$ 
and infer from Eqs. (3.26)-(3.28) that 
$$E_{\rm eq}\bigl({\tilde {\zeta}}_{t,s}(G)
{\tilde {\zeta}}_{t^{\prime},s^{\prime}}(G^{\prime})=
\bigl[2{\kappa}{\langle}a,a^{\prime}{\rangle}+$$
$${\beta}^{-1}{\gamma}_{1}{\langle}c^{(1)},c^{{\prime}(1)}
{\rangle}+2{\beta}^{-1}({\gamma}_{2}-2d^{-1}{\gamma}_{1})
{\langle}c^{(2)},c^{{\prime}(2)}{\rangle}\bigr]
{\vert}[s,t]{\cap}[s^{\prime},t^{\prime}]{\vert}.\eqno(3.29)$$
\vskip 0.2cm
This formula and Eq. (3.19) constitute our equilibrium conditions. In order to obtain 
their local properties, we consider their forms when their test functions are 
concentrated around points of ${\Omega}$. Thus, for $x_{0}{\in}{\Omega}$ and 
${\epsilon}{\in}{\bf R}_{+}$, we define the transformations 
$F{\rightarrow}F_{x_{0},{\epsilon}}$ and $G{\rightarrow}G_{x_{0},{\epsilon}}$ 
of ${\tilde {\cal D}}({\Omega})$ and ${\check {\cal D}}({\Omega})$, respectively, 
by the equations
$$F_{x_{0},{\epsilon}}={\epsilon}^{-d/2}F\bigl({\epsilon}^{-1}(x-x_{0})\bigr)
\eqno(3.30)$$
and                                                                                                                                                                             
$$G_{x_{0},{\epsilon}}={\epsilon}^{-d/2}G\bigl({\epsilon}^{-1}(x-x_{0})\bigr)
\eqno(3.31).$$
We then remark that Eqs. (3.19) and (3.29) are invariant under these transformations 
and therefore that they enjoy the local (punctual!) property that
$${\rm lim}_{{\epsilon}{\downarrow}0}E_{\rm eq}
\bigl({\xi}(F_{x_{0},{\epsilon}}){\xi}(F_{x_{0},{\epsilon}}^{\prime})\bigr)=
{\langle}F,{\pi}^{{\prime}{\prime}}({\overline {\theta}})F^{\prime}{\rangle} \ 
{\forall} \ x_{0}{\in}{\Omega}, F,F^{\prime}{\in}{\tilde {\cal 
D}}({\Omega}).\eqno(3.32)$$
and
$${\rm lim}_{{\epsilon}{\downarrow}0}
E_{\rm eq}\bigl({\tilde {\zeta}}_{t,s}(G_{x_{0},{\epsilon}})
{\tilde {\zeta}}_{t^{\prime},s^{\prime}}(G_{x_{0},{\epsilon}}^{\prime})\bigr)=
2{\kappa}{\langle}a,a^{\prime}{\rangle}+$$
$${\beta}^{-1}{\gamma}_{1}{\langle}c^{(1)},c^{{\prime}(1)}
{\rangle}+2{\beta}^{-1}({\gamma}_{2}-2d^{-1}{\gamma}_{1})
{\langle}c^{(2)},c^{{\prime}(2)}{\rangle}]
{\vert}[s,t]{\cap}[s^{\prime},t^{\prime}]{\vert}.\eqno(3.33)$$
\vskip 0.3cm
{\it Local Equilibrium Conditions.} We now assume that, even in a nonequilibrium 
steady state, the two point functions of ${\xi}$ and ${\tilde {\zeta}}$ enjoy the same 
local properties as at equilibrium. Thus, bearing in mind that ${\beta}, \ 
{\overline {\theta}}, \ {\kappa}, {\gamma}_{1}$ and ${\gamma}_{2}$, are generally 
position dependent in the nonequilibrium situation, we take the local equilibrium 
conditions to be the following ones. 
$${\rm lim}_{{\epsilon}{\downarrow}0}E
\bigl({\xi}(F_{x_{0},{\epsilon}}){\xi}(F_{x_{0},{\epsilon}}^{\prime})\bigr)=
{\langle}F,{\pi}^{{\prime}{\prime}}({\overline{\theta}}(x_{0}))F^{\prime}
{\rangle} \ {\forall} \ x_{0}{\in}{\Omega}, F,F^{\prime}{\in}
{\tilde {\cal D}}({\Omega}).\eqno(3.34)$$
and
$${\rm lim}_{{\epsilon}{\downarrow}0}
E\bigl({\tilde {\zeta}}_{t,s}(G_{x_{0},{\epsilon}})
{\tilde {\zeta}}_{t^{\prime},s^{\prime}}(G_{x_{0},{\epsilon}}^{\prime})\bigr)=
\bigl[2{\kappa}(x_{0}){\langle}a,a^{\prime}{\rangle}+$$
$${\beta}(x_{0})^{-1}{\gamma}_{1}(x_{0})
{\langle}c^{(1)},c^{{\prime}(1)}{\rangle}
+2{\beta}(x_{0})^{-1}({\gamma}_{2}(x_{0})-2d^{-1}{\gamma}_{1}(x_{0}))
{\langle}c^{(2)},c^{{\prime}(2)}{\rangle}\bigr]
{\vert}[s,t]{\cap}[s^{\prime},t^{\prime}]{\vert}.\eqno(3.35)$$
\vskip 0.3cm
The following Proposition, whose proof ensues from a trivial modification of that of 
[5, Prop. 5.3], provides an explicit formula for the bilinear form ${\Gamma}$, which 
governs the form of the two-point function for ${\tilde {\zeta}}$ according to Eq. 
(3.14).
\vskip 0.3cm
{\bf Proposition 3.6.} {\it Under the assumptions of Prop. 3.2, together with the local 
equilibrium condition (3.35), ${\Gamma}$ is given by the following formula.
$${\Gamma}(G,G^{\prime})=2{\langle}a,{\kappa}a^{\prime}{\rangle}+
{\langle}c^{(1)},{\beta}^{-1}{\gamma}_{1}c^{{\prime}(1)}{\rangle}+
2{\langle}c^{(2)},{\beta}^{-1}({\gamma}_{2}-2d^{-1}{\gamma}_{1})
c^{{\prime}(2)}{\rangle},\eqno(3.36)$$
where $a, \ a^{\prime}, \ c^{(1)}, c^{{\prime}(1)}\ c^{(2)}, \ c^{{\prime}(2)}$ are 
related to $G$ and $G^{\prime}$ according to the above specifications and now 
${\beta}^{1}, \ {\kappa}, \ {\gamma}_{1}$ are functions of position, through their 
dependence on ${\beta}$ and ${\mu}$, that act multiplicatively on 
${\check {\cal D}}({\Omega})$.}
\vskip 0.5cm
\centerline {\bf  3.4.The Macrostatistical Model.}
\vskip 0.3cm 
The stationary processes ${\xi}$ and $w$, which are connected by the integrated 
Langevin equation (3.10a), comprise our macrostatistical model. In view of Cor. (3.3) 
and Prop. (3.4), these processes are both Gaussian, and the two-point function of $w$ 
is given by Eqs. (3.15) and (3.36). To complete the formulation of the model, it 
remains for us to obtain the two-point function of the process ${\xi}$.
\vskip 0.2cm
To this end, we infer from Eq. (3.10a) that, since ${\cal L}$ is the generator of the 
semigroup $T$,
$${\xi}_{t}=T_{t-t_{0}}{\xi}_{t_{0}}+\int_{t_{0}}^{t}T_{t-s}dw_{s,t_{0}} \ 
{\forall} \ t{\geq}t_{0}$$
and hence that
$${\xi}_{t}(F)={\xi}_{t_{0}}(T_{t-t_{0}}^{\star}F)+\int_{t_{0}}^{t}
dw_{s,t_{0}}(T_{t-s}^{\star}F) \ {\forall} \ F{\in}{\tilde {\cal D}}({\Omega}), \ 
t,t_{0}
({\leq}t){\in}{\bf R}.\eqno(3.37)$$
In view of the stationarity of the ${\xi}$-process, it follows from Eq.  (3.37) and Prop. 
3.4 that the static two-point function for the field ${\xi}$ is
$$W(F,F^{\prime}):=E\bigl({\xi}(F){\xi}(F^{\prime}\bigr)=
E\bigl({\xi}(T_{t-t_{0}}^{\star}F){\xi}(T_{t-t_{0}}^{\star}F^{\prime}\bigr)+$$
$$\int_{t_{0}}^{t}ds
{\Gamma}({\nabla}T_{s}^{\star}F,{\nabla}T_{s}^{\star}F^{\prime}) \ 
{\forall} \ F,F^{\prime}{\in}{\tilde {\cal D}}({\Omega}), \ 
t,t_{0}({\leq}t){\in}{\bf R}\eqno(3.38)$$ 
On invoking the dissipative condition (2.31) and passing to the limiting form of Eq. 
(3.38) as $t_{0}{\rightarrow}-{\infty}$, we obtain the formula 
$$W(F,F^{\prime})= \int_{0}^{\infty}ds
{\Gamma}({\nabla}T_{s}^{\star}F,{\nabla}T_{s}^{\star}F^{\prime}) \ 
{\forall} \ F,F^{\prime}{\in}{\tilde {\cal D}}({\Omega}).\eqno(3.39)$$ 
We note that, as ${\cal L}^{\star}$ is the generator of the semigroup $T^{\star}$, it 
follows from this formula and Eq. (3.38) that
$$W({\cal L}^{\star}F,F^{\prime})+W(F,{\cal 
L}^{\star}F^{\prime})+{\Gamma}({\nabla}F,{\nabla}F^{\prime})=0.\eqno(3.40)$$
Moreover, in view of the stationarity of the process ${\xi}$, it follows from Eqs. 
(3.38) and (3.39) that the two-point space time correlation function is given by the 
formula
$$E\bigl({\xi}_{t}(F){\xi}_{t^{\prime}}(F^{\prime}\bigr)= \int_{0}^{\infty}ds
{\Gamma}({\nabla}T_{t-t^{\prime}+s}^{\star}F,{\nabla}T_{s}^{\star}F^{\prime}) \ 
{\forall} \ F,F^{\prime}{\in}{\tilde {\cal D}}({\Omega}), \ t,t^{\prime} 
({\leq}t){\in}{\bf R}.\eqno(3.41)$$
As the process ${\xi}$ is Gaussian, this completes our formulation of the model.
\vskip 0.5cm
\centerline {\bf 4. Long Range Correlations.}
\vskip 0.3cm	
We term the space correlations of ${\xi}$ to be of  short range, which we idealise as 
zero range in our macroscopic scaling, if the support of its static two-point function 
$W$ lies in the region 
${\lbrace}(x,x^{\prime}){\in}{\Omega}^{2}{\vert}x=x^{\prime}{\rbrace}$. Then, 
by direct analogy with the proof of  [5, Cor. 5.4], it follows from Schwartz\rq s point 
and compact support theorems [6, Ths. 35 and 26] that if the correlations are of short 
range then $W$ takes the form 
$$W(F,F^{\prime})=
{\langle}F,{\pi}^{{\prime}{\prime}}({\theta})F^{\prime}{\rangle} \ {\forall} \ 
F,F^{\prime}{\in}{\tilde {\cal D}}({\Omega}).\eqno(4.1)$$
On the other hand, we term the static space correlations of ${\xi}$ to be of long range 
if this condition is violated. Thus, in the present context, \lq long\rq\ is taken to mean 
non-zero on the employed macroscopic scale. 
\vskip 0.2cm
Our aim now is to establish that these correlations are generically of long range, i.e. 
that it is only in exceptional circumstances that the condition (4.1) is valid. To this 
end, we note that, in view of  Eqs. (2.35) and (3.40), this condition may be expresses 
in the form
$${\langle}F,{\Lambda}^{\star}F^{\prime}{\rangle}+
{\langle}F^{\prime},{\Lambda}^{\star}F{\rangle}={\Gamma}
({\nabla}F,{\nabla}F^{\prime})\eqno(4.2)$$.
Thus, the condition for long range correlations is that of the violation of Eq. (4.2) for 
some $F$ and $F^{\prime}$ in ${\tilde {\cal D}}({\Omega})$. 
\vskip 0.3cm  
{\bf Proposition 4.1.} {\it A sufficient condition for the process ${\xi}$ to have long 
range space correlations is that one of the following ones is violated. 
$$u=0\eqno(4.3)$$
and
$${\nabla}.\bigl(({\beta}^{-1}{\mu}{\kappa}_{\mu}-{\kappa}_{\beta})
{\nabla}{\beta}\bigr)=0,\eqno(4.4)$$
where ${\kappa}_{\beta}$ and ${\kappa}_{\mu}$ are the derivatives of ${\kappa}$ 
w.r.t. ${\beta}$ and ${\mu}$, respectively. Since these conditions can be satisfied 
only by certain particular forms of the space-dependent variable ${\beta}, \ {\mu}$ 
and $u$, this signifies that the space correlations of ${\xi}$ are generically of long 
range.}
\vskip 0.3cm
{\bf Remark.} It will be seen that the proof of this Proposition is based on the choice 
$F=F^{\prime}=(f,0,0)$ for the forms of the test functions, with the result that long 
range correlations prevail if either the condition (4.3) for the drift velocity or (4.4) for 
the thermal conductivity is violated. We remark here that further sufficient conditions 
for long range correlations, expressed in terms of the bulk and shear viscosities, may 
similarly be derived from other choices of the forms of $F$ and $F^{\prime}$. 
\vskip 0.3cm
{\bf Proof of Prop. 4.1.} It suffices to show that the condition (4.2) for zero range 
spatial correlations implies Eqs. (4.3) and (4.4). To this end, we choose both $F$ and 
$F^{\prime}$ to be $(f,0,0)$ and infer from Eq. (3.36) that 
$${\Gamma}({\nabla}F,{\nabla}F)=
2\int_{\Omega}dx{\kappa}({\nabla}f)^{2}.\eqno(4.5)$$ 
On the other hand, by Eqs. (2.32) and  (2.33),
$${\langle}F,{\Lambda}^{\star}F^{\prime}{\rangle}=
{\langle}{\Lambda}F,F^{\prime}{\rangle}={{\partial}\over {\partial}{\lambda}}
{\langle}{\Psi}({\overline {\theta}}+{\lambda}F),{\nabla}F^{\prime}
{\rangle}_{{\vert}{{\lambda}=0}} \ {\forall} \ F,F^{\prime}{\in}
{\tilde {\cal D}}({\Omega}).\eqno(4.6)$$ 
In order to treat this formula for the case where $F=F^{\prime}=(f,0,0)$, we define
$${\theta}_{{\lambda}f}(x):={\overline {\theta}}+{\lambda}\bigl(f(x),0,0\bigr)
\eqno(4.7)$$
On asserting the ${\theta}$-dependence of the energy current $q$ by referring to 
$q(x)$ as $q({\theta};x)-$, defining
$$q_{{\lambda}f}(x):=q({\theta}_{{\lambda}f};x)\eqno(4.8)$$
and noting that, by identification of Eq. (2.15) with Eq. (2.32), ${\Psi}:=(q,j,{\tau})$, 
we infer from Eq. (4.6) that
$${\langle}F,{\Lambda}^{\star}F{\rangle}=
\bigl[{{\partial}\over {\partial}{\lambda}}\int_{\Omega}dx
{\nabla}f(x).q_{{\lambda}f}(x)\bigr]_{{\vert}{\lambda}=0}.\eqno(4.9)$$
In order to determine the explicit form of $q_{{\lambda}f}$ in terms of the position 
dependent variables ${\beta}, \ {\mu}$ and $u$, we define their canonical 
counterparts ${\beta}_{{\lambda}f}, \ {\mu}_{{\lambda}f}$ and $u_{{\lambda}f}$, 
respectively, that correspond to ${\theta}_{{\lambda}f}$ by the version of Eq. (2.9) 
obtained by imposing the subscript ${\lambda}f$ to each of its terms. Thus 
$${\theta}_{{\lambda}f}={\beta}_{{\lambda}f}\bigl(1, \ - 
{\mu}_{{\lambda}f}+{1\over 2}u_{{\lambda}f}^{2},  \ -u_{{\lambda}f}\bigr).
\eqno(4.10)$$ 
It then follows from Eqs. (2.9) and (4.10) that
$${\beta}_{{\lambda}f}={\beta}+{\lambda}f, \ 
{\mu}_{{\lambda}f}=(1+{\lambda}{\beta}^{-1}f)^{-1}{\mu}-
{\lambda}{\beta}f(1+{\lambda}{\beta}^{-1}f)^{-2}u^{2} \ {\rm and}$$
$$u_{{\lambda}f}=(1+{\lambda}{\beta}^{-1}f)^{-1}u.\eqno(4.11)$$
Further, since the energy current, $q$, defined by Eq. (2.16), is a functional of 
${\theta}$, we express $q_{{\lambda}f}$ as the corresponding functional of 
${\theta}_{{\lambda}f}$, i.e. of ${\beta}_{{\lambda}f}, \ {\mu}_{{\lambda}f}$ and 
$u_{{\lambda}f}$. Thus,
$$q_{{\lambda}f}={\varepsilon}({\beta}_{{\lambda}f},{\mu}_{{\lambda}f})
u_{{\lambda}f}+
{1\over 2}{\rho}({\beta}_{{\lambda}f},{\mu}_{{\lambda}f})
u_{{\lambda}f}^{2}u_{\lambda}-$$
$${\gamma}_{1}({\beta}_{{\lambda}f},{\mu}_{{\lambda}f})
\bigl(Du_{{\lambda}f}-2d^{-1}{\nabla}.u_{{\lambda}f}I\bigr).u_{{\lambda}f}-
{\gamma}_{2}({\beta}_{{\lambda}f},{\mu}_{{\lambda}f})
({\nabla}.u_{{\lambda}f})u_{{\lambda}f}
+{\kappa}({\beta}_{{\lambda}f},{\mu}_{{\lambda}f})
{\nabla}{\beta}_{{\lambda}f}.\eqno(4.12)$$
Since, by Eq. (4.11),
$$\bigl({{\partial}{\beta}_{{\lambda}f}\over 
{\partial}{\lambda}}\bigr)_{{\vert}{\lambda}=0}=f; \  
\bigl({{\partial}{\mu}_{{\lambda}f}\over 
{\partial}{\lambda}}\bigr)_{{\vert}{\lambda}=0}=
-{\beta}^{-1}({\mu}+u^{2})f; \ {\rm and} \ 
\bigl({{\partial}u_{{\lambda}f}\over 
{\partial}{\lambda}}\bigr)_{{\vert}{\lambda}0}=-{\beta}^{-1}uf,\eqno(4.13)$$ 
it follows from Eqs. (4.11)-(4.13) that 
$$\bigl({{\partial}q_{{\lambda}f}\over 
{\partial}{\lambda}}\bigr)_{{\vert}{\lambda}=0}
={\kappa}{\nabla}f+{\beta}^{-1}{\gamma}_{1}
\bigl(u{\nabla}f+({\nabla}f)u\bigr)-2d^{-1}{\nabla}.(fu)I\bigr).u
+({\gamma}_{2}(u.{\nabla}f)u+$$
$$({\kappa}_{\beta}-{\beta}^{-1}{\mu}{\kappa}_{\mu})
({\nabla}{\beta})f+C({\beta},{\mu};u)f,\eqno(4.14)$$
where ${\kappa}_{\beta}$ and ${\kappa}_{\mu}$ are the derivatives of ${\kappa}$ 
w.r.t. ${\beta}$ and ${\mu}$, respectively, and $C$ is a vector valued functional of 
${\beta}, \ {\mu}$ and $u$ that vanishes when $u=0$. 
\vskip 0.2cm
It follows now from Eqs. (4.5) and (4.14) that the condition (4.2) for short range 
correlations reduces to the formula 
$$\int_{\Omega}dx{\beta}^{-1}\bigl[{\gamma}_{1}\bigl(({\nabla}f)^{2}u^{2}+
(1-2d^{-1})(u.{\nabla}f)^{2}\bigr)+{\gamma}_{2}(u.{\nabla}f)^{2}\bigr)\bigr]+$$
$$\int_{\Omega}dx{1\over 2}f^{2}{\nabla}.
\bigl(({\beta}^{-1}{\mu}{\kappa}_{\mu}-{\kappa}_{\beta})
{\nabla}{\beta}-C\bigr)=0 \ {\forall} \ f{\in}{\cal D}({\Omega}).\eqno(4.15)$$
In particular, if  we replace $f$ here by $f_{x_{0},{\epsilon}}$ where 
$x_{0}{\in}{\Omega}, \ {\epsilon}{\in}{\bf R}_{+}$ and
$$f_{x_{0},{\epsilon}}={\epsilon}^{(1-d)/2}f\bigl({\epsilon}^{-1}(x-x^{0})\bigr),
\eqno(4.16)$$
then the passage to the limit ${\epsilon}{\rightarrow}0$ annihilates the second 
integral and yields the equation
$$\int_{\Omega}dx{\beta}(x_{0})^{-1}\bigl[{\gamma}_{1}(x_{0})
\bigl(({\nabla}f(x))^{2}u(x_{0})^{2}+
(1-2d^{-1})(u(x_{0}).{\nabla}f(x))^{2}\bigr)+$$
$${\gamma}_{2}(x_{0})\bigl(u(x_{0}).{\nabla}f(x)\bigr)^{2}\bigr]
=0 \ {\forall} \ x_{0}{\in}{\Omega}, \ f{\in}{\cal D}({\Omega}).\eqno(4.17)$$
For $d{\geq}2$, this implies that the velocity field $u$ vanishes and consequently 
that the condition (4.15) reduces to that of the vanishing of the second integral of that 
equation, with $C=0$; and for $d=1, \ {\gamma}_{1}$ may be equated to zero in the 
NS equation (2.18) and therefore the same conclusion is valid. Thus, the short range 
correlation condition implies Eqs. (4.3) and (4.4), and therefore the violation of either 
of those formulae implies that the ${\xi}$ process has long range spatial correlations
\vskip 0.5cm
\centerline {\bf 5. Concluding Remarks}
\vskip 0.3cm
We have shown that the stochastic process executed by the hydrodynamical 
fluctuations of the continuum model about a nonequilibrium steady state is 
completely determined by the conditions of Onsagerian regression, local stability and 
chaoticity. This process, fully specified in Section 3.4, constitutes a mathematical 
generalisation of Landau\rq s picture of hydrodynamical fluctuations. In particular, by 
Prop. 4.1, the process generically carries long range spatial correlations. This result 
appears to be new within the framework of the Navier-Stokes equations, though it has 
been previously suggested on rather general heuristic grounds [9, 10] and proved for a 
certain classical stochastic (non-Hamiltonian) model [11]-[13]. Most notably, it marks 
an important difference between equilibrium and nonequilibrium properties of 
hydrodynamical fluctuations, as the spatial correlations of the former are of short 
(microscopic) range, except at critical points. We remark here that there is no 
corresponding qualitative difference between the time correlations of the 
hydrodynamical fluctuations about equilibrium and nonequilibrium, since the 
regression hypothesis implies that the time scales of both are the macroscopic ones of 
the Navier-Stokes flow.
\vskip 0.2cm
The treatment of hydrodynamical fluctuations in this article has been based on a 
classical macroscopic continuum model, ${\Sigma}$, of a fluid. Presumably this 
should arise from an underlying quantum mechanics, at the microscopic level, in the 
following way. One assumes that the quantum system, ${\Sigma}_{\rm qu}$, 
consists of $N$ particles of one species that is confined to the region ${\Omega}$ and 
coupled at the boundary ${\partial}{\Omega}$ to an array, ${\cal R}$, of reservoirs 
whose temperatures and chemical potentials are just those of the macrostatistical 
model. Assuming that these control variables are not the same for all the reservoirs, 
the system $({\Sigma}_{\rm qu}+{\cal R})$ will evolve, under rather general 
conditions, to a nonequilibrium steady state ${\omega}$ [14]. The hydrodynamical 
observables of ${\Sigma}_{\rm qu}$ may then be formulated  along the lines of [4, 5] 
in terms of the natural counterparts ${\hat {\phi}}_{\rm qu}$ and ${\hat {\chi}}_{\rm 
qu}$, of the random classical fields ${\hat {\phi}}$ and ${\hat {\chi}}$ of the model 
${\Sigma}$, though those quantum fields are now operator valued functions of the 
positions and momenta of the particles of the system. We denote their evolutes at time 
$t$, as governed by the dynamics of the composite $({\Sigma}_{\rm qu}+{\cal R})$ 
by 
${\hat {\phi}}_{{\rm qu};t}$ and ${\hat {\chi}}_{{\rm qu};t}$, and we assume that 
these are distributions of class ${\tilde {\cal D}}^{\prime}({\Omega})$ and 
${\check {\cal D}}^{\prime}({\Omega})$ and that their expectation values for the 
state ${\omega}$ are the classical steady state fields ${\phi}$ and ${\chi}$, 
respectively, of Section 2. We then define the quantum fluctuation fields 
${\xi}_{{\rm qu};t}$ and ${\zeta}_{{\rm qu};t,s}$ by the canonical analogues of 
Eqs. (3.2) and (3.4) and denote the smeared fields obtained by integrating them 
against test functions $F \ ({\in}{\tilde {\cal D}}({\Omega}))$ and $G \ ({\in}
{\check {\cal D}}({\Omega}))$ by ${\xi}_{{\rm qu};t}(F)$ and 
${\zeta}_{{\rm qu};t,s}(G)$, respectively. The correlation functions given by the 
expectation values, for the state ${\omega}$ of the monomials in the 
${\xi}_{{\rm qu};t}(F)$\rq s and ${\zeta}_{{\rm qu};t,s}(G)$\rq s then represent the 
quantum stochastic process [15] executed by the hydrodynamical fluctuations. 
Further, under a condition of {\it macroscopic classicality}, whereby the correlation 
functions are invariant under reordering of the constituent smeared fields, this process 
is classical. Thus, under the assumption that this condition is fulfilled, possibly up to 
corrections that are $o(1)$ w.r.t. ${\hbar}, \ k_{B}$ and microscopic relaxation times 
and attenuation lengths, the fluctuation process simulates a classical one The further 
assumptions of  Onsagerian regression, local stability and chaoticity then lead 
precisely to the classical macrostatistical one presented here. The ultimate test of the 
physical validity of that model is that its correlation functions are those of the 
hydrodynamical observables of ${\Sigma}_{\rm qu}$, up to the above microscopic 
corrections 
\vskip 0.5cm
\centerline {\bf Appendix A: Derivation of the Formula (2.36)}
\vskip 0.3cm 
We assume that, at equilibrium, $u=0$ and ${\beta}$ and ${\mu}$ are spatially 
uniform. The same is therefore true of $e, \ {\rho}, {\kappa},\ {\gamma}_{1}$ and 
${\gamma}_{2}$, since these are functions of the latter two variables. Thus, by Eq. 
(2.9a), 
$${\overline {\theta}}=({\beta},-{\beta}{\mu},0)\eqno(A.1)$$
and, as ${\phi}=(e,{\rho},{\rho}u)$, it follows from Eq. (2.33), together with the 
identification of the NS equations (2.15)-(2.19) with Eq. (2.32), that the equilibrium 
form, 
${\Lambda}_{\rm eq}$, of ${\Lambda}$ is given by the formula 
$${\Lambda}_{\rm eq}{\delta}({\beta},-{\beta}{\mu},-{\beta}u)=$$
$$\bigl(-{\varepsilon}{\nabla}.{\delta}u-{\kappa}{\Delta}{\delta}{\beta}, \ 
-{\rho}{\nabla}.{\delta}u, \ -p_{\beta}{\nabla}{\delta}{\beta}-p_{\mu}
{\nabla}{\delta}{\mu}+{\gamma}_{1}{\nabla}.(D{\delta}u-
2d^{-1}({\nabla}.{\delta}u)I)+
{\gamma}_{2}{\nabla}({\nabla}.{\delta}u)\bigr),\eqno(A.2)$$
where $p_{\beta}$ and $p_{\mu}$ are the derivatives of $p$ w.r.t. ${\beta}$ and 
${\mu}$, respectively. Since ${\phi}=(e,{\rho},{\rho}u)$, it follows from Eqs. (2.5) 
and (2.10)-(2.12) that
$$p_{\beta}=-{\beta}^{-1}({\varepsilon}-{\rho}{\mu}) \ {\rm and} \ 
p_{\mu}={\rho}.\eqno(A.3)$$
Hence, by Eqs. (A.2) and (A.3),
$${\Lambda}_{\rm eq}{\delta}{\theta}{\equiv}{\Lambda}_{eq}({\delta}{\beta},-
{\beta}{\delta}{\mu}-{\mu}{\delta}{\beta},-{\beta}{\delta}u)=$$
$$\bigl(-{\varepsilon}{\nabla}.{\delta}u-{\kappa}{\Delta}{\delta}{\beta}, \ 
-{\rho}{\nabla}.{\delta}u, \ 
{\beta}^{-1}({\varepsilon}-{\mu}{\rho}){\nabla}{\delta}{\beta}-
{\rho}{\nabla}{\delta}{\mu}+$$
$${\gamma}_{1}{\nabla}.(D{\delta}u-2d^{-1}({\nabla}.{\delta}u)I)+
{\gamma}_{2}{\nabla}({\nabla}.{\delta}u)\bigr).\eqno(A.4)$$
Equivalently, defining $({\delta}{\theta}^{(1)}, \ {\delta}{\theta}^{(2)}, \ 
{\delta}{\theta}^{(3)}):={\delta}{\theta}=({\delta}{\beta}, \ -{\beta}{\delta}{\mu}-
{\mu}{\delta}{\beta}, \ –{\beta}{\delta}u)$, 
$${\Lambda}_{\rm eq}{\delta}{\theta}=
{\beta}^{-1}\bigl({\varepsilon}{\nabla}.{\delta}{\theta}^{(3)}-
{\beta}{\kappa}{\Delta}{\delta}{\theta}^{(1)}, \ 
{\rho}{\nabla}.{\delta}{\theta}^{(2)}, \ $$ 
$${\varepsilon}{\nabla}{\delta}{\theta}^{(1)}+{\rho}{\nabla}{\delta}{\theta}^{(2)}
+{\gamma}_{1}{\nabla}.(D{\delta}{\theta}^{(3)}-
2d^{-1}{\nabla}.{\delta}{\theta}^{(3)})+{\gamma}_{2}
{\nabla}({\nabla}.{\delta}{\theta}^{(3)}\bigr).\eqno(A.5)$$
Further, the dual, ${\Lambda}_{\rm eq}^{\star}$, of ${\Lambda}_{\rm eq}$ is 
defined by the identity
$$\bigl({\delta}{\theta}, \ {\Lambda}_{\rm eq}^{\star}F\bigr){\equiv}
\bigl({\Lambda}_{\rm eq}{\delta}{\theta}, \ F\bigr) \ {\forall} \ 
F{\in}{\tilde {\cal D}}({\Omega}).\eqno(A.6),$$
The formula (2.36) follows immediately from Eqs. (A.5) and (A.6). 
\vskip 0.5cm
\centerline {\bf Appendix B. Proof of Lemma 3.5.} 
\vskip 0.3cm
By the standard relationship between distributions and the corresponding generalised 
functions,
$${\tilde {\zeta}}_{t,s}^{(1)}({\nabla}f)=-\int_{\Omega}dxf(x)
{{\partial}\over {\partial}x_{i}}{\tilde {\zeta}}_{t,s;i}^{(1)}(x) \ {\forall} \ 
f{\in}{\cal D}({\Omega})\eqno(B.1)$$
and
$${\tilde {\zeta}}_{t,s}^{(3)}({\nabla}h)=-\int_{\Omega}dxh_{i}(x)
{{\partial}\over {\partial}x_{j}}{\tilde {\zeta}}_{t,s;i,j}^{(3)}(x)  \ 
{\forall} \ h{\in}{\cal D}_{T}({\Omega}).\eqno(B.2)$$
On combining these formulae with  Eqs. (2.19), (3.21) and (3.22), we see that
$$\int_{{\Omega}^{2}}dxdx^{\prime}f(x)f^{\prime}(x^{\prime})
{{\partial}^{2}\over {\partial}x_{i}{\partial}x_{j}^{\prime}}
E_{eq}\bigl({\tilde {\zeta}}_{t,s;i}^{(1)}(x)
{\tilde {\zeta}}_{t^{\prime},s^{\prime};j}(x^{\prime})\bigr)=$$
$$ 2{\kappa}\int_{{\Omega}^{2}}dxdx^{\prime}
f(x)f^{\prime}(x^{\prime}){{\partial}^{2}\over 
{\partial}x_{i}{\partial}x_{j}^{\prime}}{\delta}(x-x^{\prime}){\times}$$
$${\vert}[s,t]{\cap}[s^{\prime},t^{\prime}{\vert}
 \ {\forall} \ f,f^{\prime}{\in}{\cal D}({\Omega}) 
  \ t,s,t^{\prime},s^{\prime}{\in}{\bf R}.\eqno(B.3)$$
and 
$$\int_{{\Omega}^{2}}dxdx^{\prime}h_{i}(x)h_{j}(x^{\prime})
{{\partial}^{2}\over {\partial}x_{l}{\partial}x_{m}^{\prime}}
E_{\rm eq}\bigl({\tilde {\zeta}}_{t,s;il}^{(3)}(x)
{\tilde {\zeta}}_{t^{\prime},s^{\prime};jm}^{(3)}(x^{\prime})\bigr)= $$
$$2{\beta}^{-1}\int_{{\Omega}^{2}}dxdx^{\prime}h_{i}(x)h_{j}(x^{\prime})
{{\partial}^{2}\over {\partial}x_{l}{\partial}x_{m}^{\prime}}
\bigl({\gamma}_{1}({\delta}_{ij}{\delta}_{lm}+{\delta}_{im}{\delta}_{jl}
-2d^{-1}{\delta}_{il}{\delta}_{jm})
+{\gamma}_{2}{\delta}_{il}{\delta}_{jm}\bigr){\delta}(x-x^{\prime}){\times}$$
$${\vert}[s,t]{\cap}[s^{\prime},t^{\prime}]{\vert}
 \ {\forall} \ h,h^{\prime}
{\in}{\cal D}_{V}({\Omega}), \ t,s,t^{\prime},s^{\prime}{\in}{\bf R}.
\eqno(B.4)$$
Eqs. (B.3) and (B.4) signify that
$${{\partial}^{2}\over {\partial}x_{i}{\partial}x_{j}^{\prime}}
\bigl[E_{\rm eq}\bigl({\tilde {\zeta}}_{t,s;i}^{(1)}(x)
{\tilde {\zeta}}_{t^{\prime},s^{\prime};j}(x^{\prime})\bigr)-
2{\kappa}{\delta}(xx^{\prime}){\vert}[s,t]{\cap}[s^{\prime},t^{\prime}{\vert}\bigr]
=0\eqno(B.5)$$
and
$${{\partial}^{2}\over {\partial}x_{l}{\partial}x_{m}^{\prime}}
\bigl[E_{\rm eq}\bigl({\tilde {\zeta}}_{t,s;il}^{(3)}(x)
{\tilde {\zeta}}_{t^{\prime},s^{\prime};jm}^{(3)}(x^{\prime})\bigr)-$$
$$2{\beta}^{-1}\bigl({\gamma}_{1}({\delta}_{ij}{\delta}_{lm}+{\delta}_{im}
{\delta}_{jl}-2d^{-1}{\delta}_{il}{\delta}_{jm})+({\gamma}_{2}
{\delta}_{il}{\delta}_{jm}\bigr){\delta}(x-x^{\prime}){\times}$$
$${\vert}[s,t]{\cap}[s^{\prime},t^{\prime}]{\vert}\bigr]=0.\eqno(B.6)$$
\vskip 0.2cm 
We now invoke our assumptions that the tensor ${\hat {\tau}}$, and hence ${\tilde 
{\zeta}}^{(3)}$, is symmetric, that the equilibrium two-point functions of ${\tilde 
{\zeta}}^{(1)}$ and ${\tilde {\zeta}}^{(3)}$ are translationally and rotationally 
invariant and that they are of zero range. It then follows that these functions take the 
following forms.
$$E_{\rm eq}\bigl({\tilde {\zeta}}_{t,s;i}^{(1)}(x){\tilde 
{\zeta}}_{t^{\prime},s^{\prime};j}(x^{\prime})\bigr)=A(x-x^{\prime}){\delta}_{ij}
\eqno(B.7)$$
and
$$E_{\rm eq}\bigl({\tilde {\zeta}}_{t,s;il}^{(3)}(x)
{\tilde {\zeta}}_{t^{\prime},s^{\prime};jm}^{(3)}(x^{\prime})\bigr)=
B(x-x^{\prime})({\delta}_{ij}{\delta}_{lm}+{\delta}_{im}{\delta}_{lj})+
C(x-x^{\prime}){\delta}_{il}{\delta}_{jm},\eqno(B.8)$$
where $A, B$ and $C$ are generalised functions on ${\Omega}$ that depend on 
$t,s,t^{\prime}$ and $s^{\prime}$. It follows from Eqs. (B.5) and (B.7) that
$${\Delta}\bigl[A(x-x^{\prime})-2{\kappa}{\delta}(x-x^{\prime})
{\vert}[s,t]{\cap}[s^{\prime},t^{\prime}]{\vert}\bigr]=0\eqno(B.9)$$
and from Eqs. (B.6) and (B.8) that
$${\Delta}\bigl[B(x-x^{\prime})-2{\beta}^{-1}{\gamma}_{1}
{\delta}(x-x^{\prime}){\vert}[s,t]{\cap}[s^{\prime},t^{\prime}]{\vert}\bigr]
=0\eqno(B.10)$$
and
$${\Delta}\bigl[C(x-x^{\prime})-
2{\beta}^{-1}({\gamma}_{2}-2d^{-1}{\gamma}_{1}){\delta}(x-x^{\prime})
{\vert}[s,t]{\cap}[s^{\prime},t^{\prime}]{\vert}\bigr]=0.\eqno(B.11)$$
In view of Schwartz\rq s point support theorem [6, Th. 35], these last three equations 
signify that
$$A(x-x^{\prime})=2{\kappa}{\delta}(x-x^{\prime})
{\vert}[s,t]{\cap}[s^{\prime},t^{\prime}]{\vert},\eqno(B.12)$$
$$B(x-x^{\prime})=2{\beta}^{-1}{\gamma}_{1}{\delta}(x-x^{\prime})
{\vert}[s,t]{\cap}[s^{\prime},t^{\prime}]{\vert}\eqno(B.13)$$
and
$$C(x-x^{\prime})=2{\beta}^{-1}({\gamma}_{2}-2d^{-1}{\gamma}_{1})
{\delta}(x-x^{\prime}){\vert}[s,t]{\cap}[s^{\prime},t^{\prime}]{\vert}.
\eqno(B.14)$$
Eq. (3.25) now follows from Eqs. (B.7) and (B.12); and Eq. (3.26) follows from Eqs. 
(B.8), (B.13) and (B.14). Finally, in view of the observation following Eq. (3.24), a 
parallel treatment of the other components of the two-point functions of ${\tilde 
{\zeta}}$ reveals that they all vanish.
\vskip 0.5cm
\centerline {\bf References}
\vskip 0.3cm\noindent
[1] A. Einstein: Ann. Phys. {\bf 11} (1903), 170; {\bf 17} (1905), 549
\vskip 0.2cm\noindent
[2] L. Onsager: Phys. Rev. {\bf 37} (1931), 405; {\bf 38} (1931), 2265
\vskip 0.2cm\noindent
[3] L. D. Landau and E. M. Lifschitz: {\it Fluid Mechanics}, Pergamon, Oxford, 1984
\vskip 0.2cm\noindent
[4] G. L. Sewell: Lett. Math. Phys. {\bf 68} (2004), 53
\vskip 0.2cm\noindent
[5] G. L. Sewell: Rev. Math. Phys. {\bf 17} (2005), 977
\vskip 0.2cm\noindent
[6] L. Schwartz: {\it Theorie des distributions}, Hermann, Paris, 1998
\vskip 0.2cm\noindent
[7] G. L. Sewell: J. Phys. A {\bf 41} (2008), 382003
\vskip 0.2cm\noindent
[8] R. F. Streater and A. S. Wightman: {\it PCT, Spin and Statistics, and All That}, 
Benjamin, New York, 1964
\vskip 0.2cm\noindent
[9] G. Grinstein, D. H. Lee and S. Sachdev: Phys. Rev. Lett. {\bf 64} (1990), 1927
\vskip 0.2cm\noindent
[10] J. R. Dorfman, T. R. Kirkpatrick and J. V. Sengers:  Ann. Rev. Chem. Phys. {\bf 
45} (1994), 213
\vskip 0.2cm\noindent
[11] H. Spohn: J. Phys. A {\bf 16} (1983), 4275
\vskip 0.2cm\noindent
[12] B. Derrida, J. L. Lebowitz and E. R. Speer: J. Stat. Phys. {\bf 107} (2002), 599
\vskip 0.2cm\noindent
[13] L. Bertini, A. de Sole, D. Gabrielli, G. Jona-Lasinio and C. Landim: J. Stat. Phys. 
{\bf 107} (2002), 635
\vskip 0.2cm\noindent
[14] D. Ruelle: J. Stat. Phys. {\bf 98} (2000), 57
\vskip 0.2cm\noindent
[15] L. Accardi, A. Frigerio and J. T. Lewis: Publ. Res. Inst. Math. Sci. {\bf 18} 
(1982), 97 
\end